\pdfoutput=1
\documentclass[journal]{IEEEtran}
\usepackage[utf8]{inputenc}
\usepackage{graphicx}
\usepackage{cite}
\usepackage{url}
\usepackage{color}
\usepackage{amsmath}
\usepackage{amssymb}
\usepackage{multirow} 
\usepackage{makecell}
\usepackage{booktabs}
\usepackage{bm}
\usepackage{algorithm} 
\usepackage{algorithmicx} 
\usepackage{algpseudocode}
\usepackage{ragged2e}
\usepackage{amssymb}
\usepackage{subcaption}

\title{Multi-Scale Feature Fusion Transformer Network for End-to-End Single Channel Speech Separation}
\author{Yinhao Xu, Jian Zhou, Liang Tao, and Hon Keung Kwan, (IEEE SENIOR MEMBER)}
\date{June 2022}

\begin{document}

\maketitle

\begin{abstract}
Recently studies on time-domain audio separation networks (TasNets) have made a great stride in speech separation. One of the most representative TasNets is a network with a dual-path segmentation approach. However, the original model called DPRNN used a fixed feature dimension and unchanged segment size throughout all layers of the network. In this paper, we propose a multi-scale feature fusion transformer network (MSFFT-Net) based on the conventional dual-path structure for single-channel speech separation. Unlike the conventional dual-path structure where only one processing path exists, adopting several iterative blocks with alternative intra-chunk and inter-chunk operations to capture local and global context information, the proposed MSFFT-Net has multiple parallel processing paths where the feature information can be exchanged between multiple parallel processing paths. Experiments show that our proposed networks based on multi-scale feature fusion structure have achieved better results than the original dual-path model on the benchmark dataset-WSJ0-2mix, where the SI-SNRi score of MSFFT-3P is 20.7dB (1.47\% improvement), and MSFFT-2P is 21.0dB (3.45\% improvement), which achieves SOTA on WSJ0-2mix without any data augmentation method.
\end{abstract}


\section{Introduction}
\label{sec:introduction}
\IEEEPARstart{S}{peech} separation is a long-standing fundamental and classical problem for speech signal processing tasks such as automatic speech recognition, speech emotion recognition and speech conversion \cite{Cherry1953, Haykin2005TheCP}. The goal of speech separation is to single out the voice of each speaker in the mixed speech with multiple speakers talking at about the same time \cite{Huang2014DeepLF, Weninger2014DiscriminativelyTR}. In this paper, we mainly focus on the single-channel speech separation task where the mixed speech signals are recorded with a single microphone.

Performing speech separation in the time-frequency (T-F) domain has been developed for decades years and remains the most commonly used methodology \cite{Wang2018SupervisedSS}. The goal of the time-frequency domain approach is to learn a T-F mask from a mixture signal. A T-F mask is essentially a weighting function (mask) estimated for each source and would be used to reconstruct the individual sources \cite{Liu2019DivideAC}. 

Typically, in the time-frequency (T-F) domain, the spectrogram of the mixture speech is obtained using the short-time Fourier transform (STFT). For each individual source, a mask is estimated from the spectrogram using statistical method or machine learning method, which is then used to multiply each T-F unit of the mixture spectrogram to recover each individual source. For instance, in the ideal binary mask (IBM) approach \cite{Wang2005OnIB}, if the signal-to-noise ratio (SNR) of a time-frequency unit in the spectrogram of $y(t)$ exceeds a predefined threshold, the mask value of the time-frequency unit is set to 1, otherwise to 0. In order to better preserve information of the target voice, the Ideal ratio mask (IRM) approach is proposed where the value of the weighting function can be any scalar ranging from 0 to 1 \cite{Zhang2016ADE, Huang2015JointOO}. Both the IBM and IRM approaches discard the phase information and adopt only the amplitude spectrum for estimating the weighting function. To tackle this issue, the complex ideal ratio mask (cIRM) is proposed and obtain a better speech separation result than the IBM and IRM methods \cite{Williamson2016ComplexRM}. 

%Although the T-F domain methodology has already made considerable progress in the field of speech separation, it still %exists a crucial problem: phase mapping. Although several methods have been proposed to deal with this problem using %phase information such as phase-sensitive mask (PSM) \cite{} and cIRM \cite{}, there exist limitation to accuracy of %estimated mask. For example, Luo et al \cite{} note that converting signals from time domain to frequency domain using %STFT leading to the difficulty of dealing with phase.

%\highlight(give some examples here by expanding some related references such as \cite{Williamson2016ComplexRM} and \cite{Hershey2016DeepCD, Isik2016SingleChannelMS, Fan2019DiscriminativeLF} )

%The implementation methods mainly include ideal binary mask (IBM), ideal ratio mask %(IRM)\cite{Zhang2016ADE, Huang2015JointOO}, complex ideal ratio mask %(cIRM)\cite{Williamson2016ComplexRM} and deep clustering\cite{Hershey2016DeepCD, %Isik2016SingleChannelMS, Fan2019DiscriminativeLF}.

Although the T-F mask based approaches achieve great progress for speech separation, there still exist some issues. Firstly, the time-frequency domain approaches use high-level acoustic feature to perform mask estimation. The lost of fine grain feature information may decrease the mask estimation accuracy, making it not be the best choice for speech separation. Secondly, the phase information of the mixture source, which is useful to reconstruct the separated signal, is only used in the training stage rather than reconstructing stage, leading to a limitation on the accuracy of the reconstructed signal. Although phase reconstruction methods are used to solve this problem, such as cIRM, there are still limitations to the performance of these methods. Thirdly, performing effective speech separation in the time-frequency domain requires a high-resolution time-frequency representation of the mixed signal, which is very time-consuming due to the high computational cost, leading to an upper bound on its applicability for real-time, low-latency applications deployed on telecommunication devices. 

Recently, the community devotes more attention to an alternative way of conducting speech separation directly in time-domain \cite{Luo2018TaSNetTA,Luo2019ConvTasNetSI,Kadolu2020AnES,FurcaNeXt,Luo2020DualPathRE,Chen2020DualPathTN,Nachmani2020VoiceSW,Zhao2020MultiScaleGT,Zeghidour2021WavesplitES,Subakan2021AttentionIA,speechbrain,Lam2021EffectiveLT,Lam2021SandglassetAL}. For instance, \cite{Luo2018TaSNetTA} proposed a time-domain audio separation network (Tasnet) to estimate the mask. In TasNet, the mixture waveform is modeled with a convolutional encoder-decoder architecture, which consists of an encoder with a non-negativity constraint on its output and a linear decoder for inverting the encoder output back to the sound waveform. Tasnet takes advantage of removing the frequency decomposition step which is required in time-frequency domain and reducing the separation problem to source masks estimation from encoder outputs which is then synthesized by the decoder. Compared with time-frequency domain based method, the Tasnet decreases the computational cost of speech separation, and significantly reduces the minimum required latency of the output, making it be more suitable for applications where low-power, real-time implementation is desirable such as in hearable and telecommunication devices \cite{Luo2019ConvTasNetSI}.

%In the past few years, a majority of research of speech separation has focused on %TasNet and proposed various TasNets in order: the LSTM-TasNet\cite{Luo2018TaSNetTA}, %the Conv-TasNet\cite{Luo2019ConvTasNetSI, Kadolu2020AnES}, the %FurcaNeXt\cite{FurcaNeXt}, the DPRNN\cite{Luo2020DualPathRE}, the %DPTNet\cite{Chen2020DualPathTN}, the gated DPRNN\cite{Nachmani2020VoiceSW}, the %MSGT-Tasnet\cite{Zhao2020MultiScaleGT}, the Wavesplit\cite{Zeghidour2021WavesplitES}, %the Sepformer\cite{Subakan2021AttentionIA, speechbrain}, the globally attentive %locally recurrent network (GALR)\cite{Lam2021EffectiveLT}, and the %Sandglasset\cite{Lam2021SandglassetAL}.

Unlike the time-frequency domain approaches, the time-domain separation systems often receive input sequences consisting of a huge number of time steps, which introduces challenges for modeling extremely long sequences. To tackle this issue, Luo \textit{et al.} proposed a dual-path recurrent neural network (DPRNN) instead of 1-D CNN in the TasNet \cite{Luo2020DualPathRE}. Specifically  DPRNN splits the input sequence into shorter chunks and interleave two RNNs, an intra-chunk RNN and an inter-chunk RNN, for local and global modeling, respectively. In a DPRNN block, the intra-chunk RNN first processes the local chunks independently, and then the inter-chunk RNN aggregates the information from all the chunks to perform utterance-level processing. Later, Chen \textit{et al.} propose a dual-path transformer network (DPTNet) for end-to-end monaural speech separation. DPTNet integrates a recurrent neural network into original transformer and embed it into a dual-path network \cite{Chen2020DualPathTN}.

However, both \cite{Luo2020DualPathRE} and \cite{Chen2020DualPathTN} use a fixed feature dimension and fixed segment size throughout all layers of the network for long sequences modeling. Although an intra-chunk RNN and an inter-chunk RNN are proposed for local and global modeling, respectively, global modeling ability is still limited. Firstly, as the intra and inter iteration processing depth increases, more and more useful context information is lost. In this case, further increasing the number of separation blocks has no significant performance improvements. Secondly, increasing the number of separation blocks may also cause the gradient vanishing problem, result in the separation performance decrease of these approaches.

In this paper, we propose a multi-scale feature fusion transformer network (MSFFT-Net) based on a novel parallel-path structure for end-to-end single-channel speech separation. Unlike the conventional dual-path structure where only one processing path exists, adopting several iterative blocks with alternative intra-chunk and inter-chunk operations to capture local and global context information, the proposed MSFFT-Net has multiple parallel processing paths where the feature information can be exchanged between adjacent parallel processing paths. We argue that the feature obtained in different processing path has different information scale. Specifically, the lower processing path focuses on high-level information processing while the upper processing path focuses on low-level information processing. The information exchange in different paths can keep the feature resolution unchanged, regardless of the processing depth. The output of the parallel path are fused together to make the extracted feature obtaining more useful global and local context information of individual source.

In order to confirm the effectiveness of the proposed MSFFT-Net, two types of MSFFT-Net, i.e., the MSFFT-2P and the MSFFT-3P are used to perform speech separation. We conducted various experiments on the benchmark dataset-WSJ0-2mix and compared the proposed MSFFT-Net with conventional dual-path based approaches such as DPRNN \cite{Luo2020DualPathRE}, DPTNet\cite{Chen2020DualPathTN} and Sepformer\cite{Subakan2021AttentionIA, speechbrain}. Experimental results show that the SI-SNRi of MSFFT-Net with 3 paths is 0.3dB higher (1.47\% improvement) than that of the Sepformer approach, and the SI-SNRi score of MSFFT-Net with 2 paths is 0.7dB higher (3.45\% improvement) than that of the Sepformer approach on the benchmark dataset-WSJ0-2mix. The proposed multi-scale feature fusion structures also outperform the original dual-path model such as DPRNN and DPTNet, indicating the fairly generic of the proposed MSFFT-Net.

In summary, the major contributions of our work are three-fold as follows. 
\begin{enumerate}
    \item A novel multi-scale speech separation network is proposed to capture the features of different scales and enrich the feature information. High-level feature and low-level feature are simultaneously processed through different processing paths, each of which in depth consists of intra- and inter chunk operation iteratively and alternatively.
    \item High-level feature and low-level feature are exchanged between different processing path using upsampling and downsampling operation, aiming to keep high feature resolution in different processing path, regardless the processing depth of the processing path. The outputs of parallel paths are fused together, enabling information from speech sequences at different scales could dynamically interact. In addition, the parallel processing path of the proposed MSFFT-Net can be trained in parallel, which would reduce the computation cost, therefore improving the efficiency of the MSFFT-Net model. 
    \item Experimental results show the proposed MSFFT-Net gains new state-of-the-art speech separation performance. We also argue that the proposed multi-scale feature fusion methodology can be generalized to other speech separation method such as the method proposed in \cite{Luo2020DualPathRE} and \cite{Chen2020DualPathTN} to obtaining significant separation performance improvement. 
\end{enumerate}

The rest of the paper is organized as follows. Related work is given in Section~\ref{sec:rw}. The proposed network is detailed in Section~\ref{sec:model design}. We devote Section~\ref{sec:experiments} and Section~\ref{sec:results} for experiments. Discussion and conclusion are given in Section~\ref{sec:discussion} and Section~\ref{sec:conclusion} respectively.

%Thirdly. We confirm the effectiveness and generality of the multi-scale feature fusion %approach by related experiments. Experiments show that our proposed networks based on %MSFF structure have achieved better results than the original dual-path models on the %benchmark dataset-WSJ0-2mix, where the SI-SNRi score of MSFFT-Net(3L) is 20.7dB (1.47\% %improvement), and MSFFT-Net(2L) is 21.0dB (3.96\% improvement).

\section{Related Work}
\label{sec:rw}
\subsection{TasNet}
\label{sec:tasnet}

For the single-channel speech separation task, the mixture speech $y(t) \in \mathbb{R}^{1 \times T}$ can be formulated as,
\begin{equation}
    \label{eq1}
    y(t)=\sum_{i=1}^{C}{s_i(t)}
\end{equation}
where $s_i(t) \in \mathbb{R}^{1 \times T}$ denote the source speech of the $i$-th speaker. The aim of single-channel speech separation is to estimate $C$ sources $s_1(t),s_2(t),..., s_C(t) \in \mathbb{R}^{1 \times T}$ from the mixture $y(t) \in \mathbb{R}^{1 \times T}$.

In the time-domain audio separation framework, the input mixture $y(t)$ is divided into $\hat{T}$ overlapping segments of length $L$, represented by $x_k \in \mathbb{R}^{1 \times L}$ which is then transformed into a $M$ dimensional representation $\mathbb{w} \in \mathbb{R}^{1 \times M}$ via a 1-D convolution operation. 
\begin{figure}[htb]
    \centering
    \includegraphics[scale=0.3]{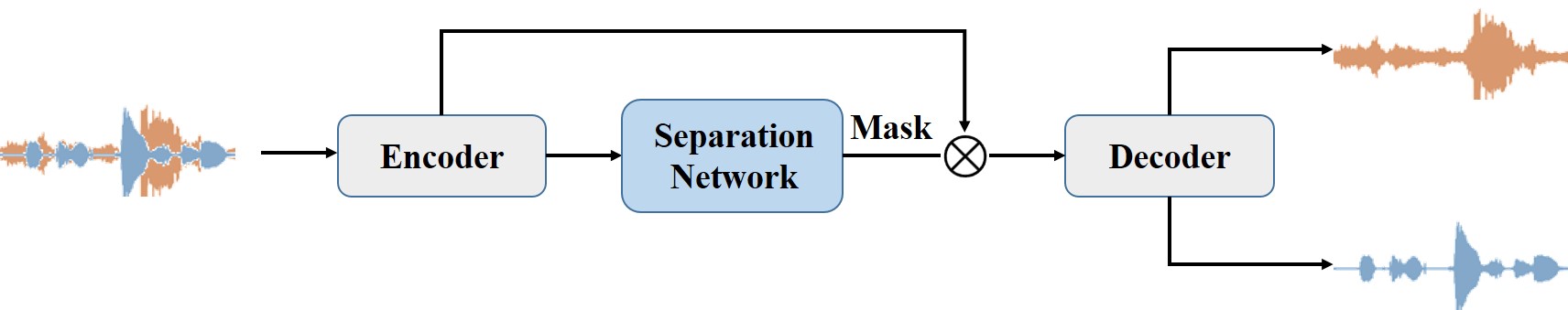}
    \caption{The system flowchart of Tasnet for speech separation.}
    \label{fig:1}
\end{figure}

Fig.\ref{fig:1} plots the basic flowchart of the time-domain audio separation framework. The encoder maps each segment $x_k$ of the mixture waveform to a high-dimensional representation which can be formulated as,
\begin{equation}
    \label{eq2}
    \tilde{X}=H(Conv1D(\mathbf{x;U}))
\end{equation}
where $\mathbf{U} \in \mathbb{R}^{E\times L} $ refers to $E$ filters with length $L$ each. $H$ denotes an optional nonlinear function such as the Relu. $ Conv1D (.) $ denotes the 1D convolution operation. Noting that the directly use of $\tilde{X}$ would lead to excessive computation cost due to its relative large dimension, we linearly map $\tilde{X}$ into $ D $-dimension features $ Y=W\tilde{X} $, where $ W\in R^{D\times E} $ and $ D $ should be smaller than $ E $.

% 2022-0808 21:36
%The encoder takes in the time-domain waveform of mixture $x\in R^T $ as signal input, then we can %divide it into L 50\%-overlapping frames, denoted by $X=[x_1, x_2,...,x_L ]\in R^{M\times L}$, %where M is a hyper parameter referred to window length. 

Unlike traditional time-frequency signal methodology, Tasnet replace STFT with a 1D-gated convolutional layer to transform each speech frame $x_i$ into a $E$-D feature vector $\tilde{x_i}$ such that we can obtain converted frames $\tilde{X}\in \mathbb{R} ^{E\times L}$.

The separation network shown in Fig.\ref{fig:2} estimates $C$ masks $m_i \in \mathbb{R}^{1 \times M}, i=1,\ldots,C$ where $C$ is the number of target sources in the mixture and $\sum m_i = 1$ with each $m_i \in [0,1]$. each individual source can be reconstructed by,
\begin{equation}
    \label{eq3}
    \mathbf{d_i}=w \odot m_i
\end{equation}
where $\odot$ denotes element-wise multiplication.

The decoder reconstructs the source waveforms from the masked features by
\begin{equation}
    \label{eq4}
    \mathbf{s_i}=d_i v.
\end{equation}

\subsection{Dual-path Processing}
\label{sec:dual-path}
\begin{figure}[htb]
    \centering
    \includegraphics[scale=0.3]{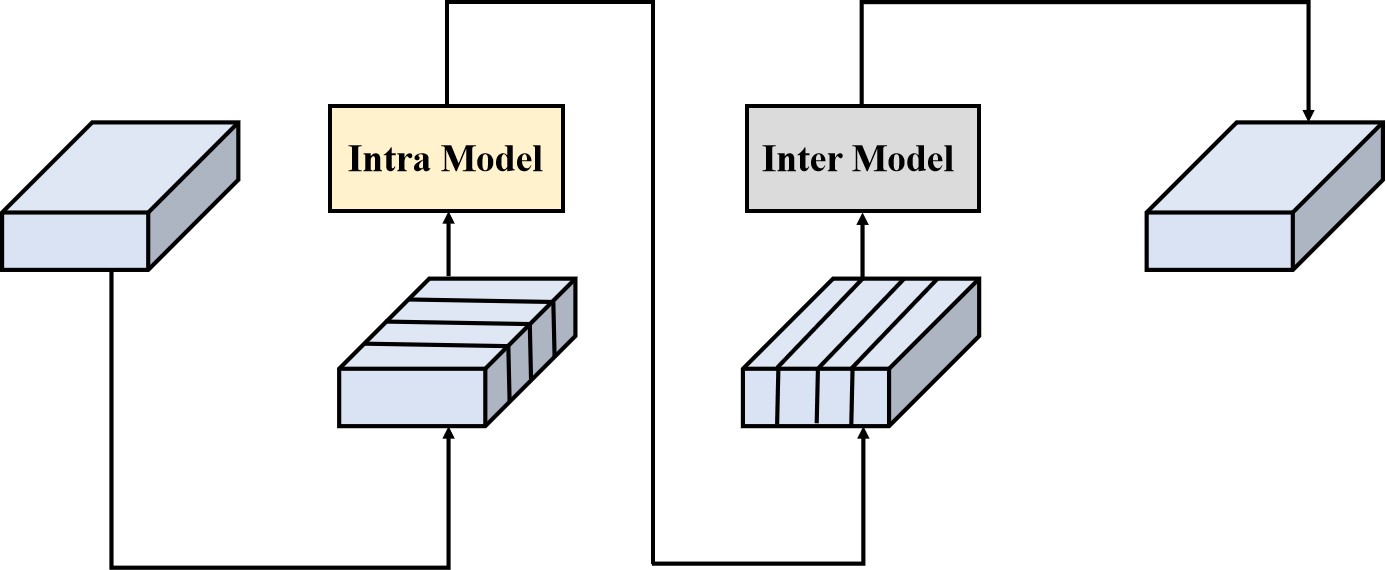}
    \caption{The overall pipeline of dual-path approach.}
    \label{fig:2}
\end{figure}

The dual-path processing is originally proposed to efficiently perform extremely long sequence modeling for time-domain single-channel speech separation. To this end, dual-path approach splits the long sequential input into smaller chunks and applies intra- and inter-model operations iteratively and alternately. Specifically, a dual-path approach consists of three processing stages: segmentation, block processing, and overlap-add. Fig.~\ref{fig:2} shows the general flowchart of dual-path processing. 

Given a long sequential input $\mathbf{W} \in \mathbb{R}^{N\times L}$ where $N$ denotes the feature dimension and $L$ denotes the number of time-steps, the segmentation stage splits $\mathbf{W}$ into $S$ overlapped chunks of length $K$ and hop size $P$ and concatenates all the chunks into a 3-D tensor $\mathbf{Z}\in R^{N\times K\times S}$.

The block processing stage consists of $M$ blocks, each of which contains two sub-modules corresponding to intra- and inter-chunk processing, respectively. Th the intra-model is applied to each independent chunk of $\mathbf{Z}$ to capture the local information and the short-term dependencies within each chunk while the inter-model is then applied to aggregate the global information of all chunks and model the long-term dependencies across the chunks. The whole process can be described as follows,
\begin{equation}
    \label{eq5}
    \mathbf{T}_{i+1}=f_{inter}(f_{intra}(\mathbf{T_{i}}), i=0,1,\ldots,M-1
\end{equation}
where $\mathbf{T}_{0}=\mathbf{Z}$, $M$ denotes the total number of blocks, $f_{inter}(\cdot)$ and $f_{intra}(\cdot)$ denote inter-model and intra-model operation respectively.

Based on the dual-path approach, a dual-path recurrent neural network (DPRNN) was firstly proposed in \cite{Luo2020DualPathRE} where a BiLSTM is employed to realize intra-model and inter-model. Inspired by dual-path RNN, Chen \textit{et al.} proposed DPTNet\cite{Chen2020DualPathTN}, which replaced the first linear layer in the original transformer with a RNN structure to capture the positional information of speech sequences. With context-aware information, DPTNet is able to effectively model very long speech sequences with relatively competitive performance. However, DPTNet has an excessively computation complexity, making it difficult to be directly applied to real application. To cope with this problem, Lam \textit{et al.} proposed a globally attentive locally recurrent network (GALR)\cite{Lam2021EffectiveLT} where BLSTM is applied to intra-chunk and self-attention network (SAN) is applied to inter-chunk. In addition, Cem Subakan \textit{et al.} proposed Sepformer\cite{Subakan2021AttentionIA}, which applies transformer encoder layer to intra-model and inter-model and achieves excellent results \cite{Hershey2016DeepCD}.

\section{The Proposed MSFF-Net for Speech Separation}
\label{sec:model design}
\subsection{The Overall Architecture of The Proposed MSFFT-Net}
\begin{figure*}[ht]
    \centering
    \includegraphics[scale=0.38]{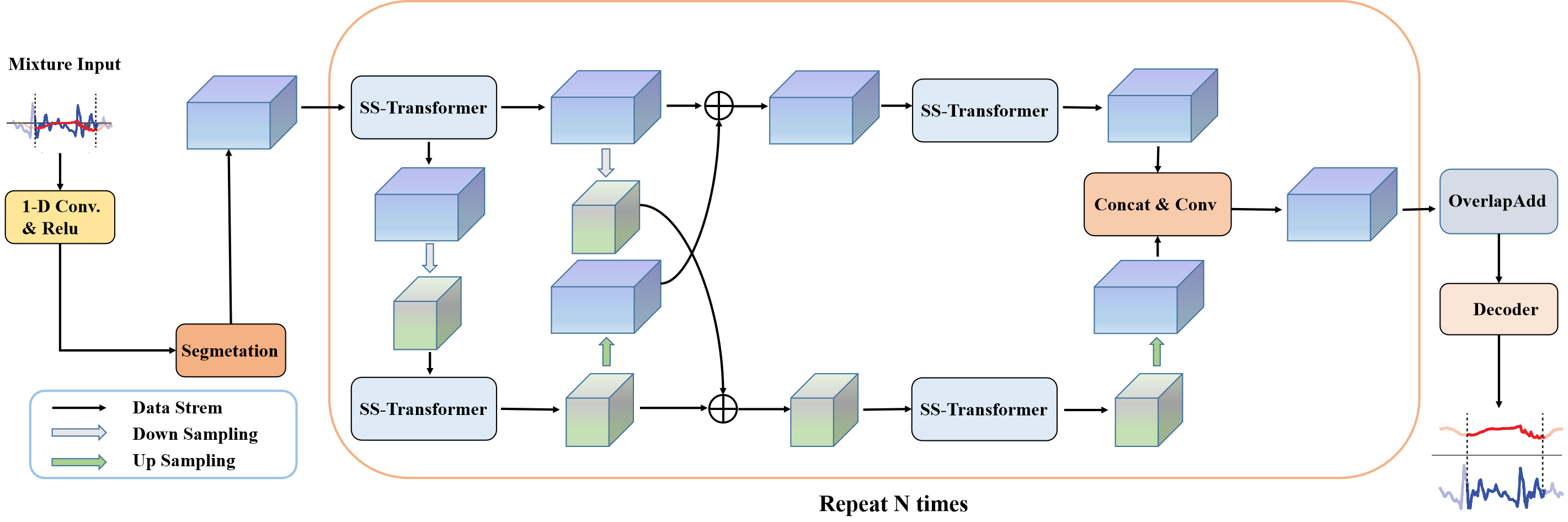}
    \caption{The architecture of MSFFT-Net for speech separation.}
    \label{fig:3}
\end{figure*}

As shown in Fig.~\ref{fig:3}, the proposed MSFFT-Net consists of encoder, separation network and decoder, which is similar to that of the DPRNN proposed in \cite{Luo2020DualPathRE}. The input of the MSFFT-Net is a mixture of speech with the length of $N$. The encoder comprises encoding and segmentation block. The waveform of mixed speech is converted into a 2-D feature map $\mathbf{W}\in \mathbb{R}^{N\times L}$ via 1-D convolutional operation in the encoding module. The segmentation module then splits $\mathbf{W}$ into $S$ overlapping segments with each length of which is $K$ and hop size $P$. The first and last segments are zero-padded so that each sample in $\mathbf{W}$ appears and only appears in $K/P$ segments. All segments are finally concatenated to be a 3-D tensor $\mathbf{Z}\in \mathbb{R}^{N\times K\times S}$. 

\label{sec:architecture}
% 2022-09-27 22:06
% 2022-10-06 15:21
% 上面的处理通路就是常规的intra-inter双路径通路，但是这样迭代下去，虽然可以建立更加精确的mask，但是一来这样会导致特征的进一步缺失，二来，过深的网络结构可能会导致梯度消失。在DPRNN提出后就有人做过通过单纯增加迭代次数的方式来测试效果，实验结果表明当迭代次数超过8次之后，效果就提升不了了。而我的思路是采用下采样-上采样-特征融合的过程，首先下采样是为了获取更粗粒度的特征，而经过上采样-特征融合之后，可以捕获到不同粒度的特征，进一步丰富特征信息，这样一来经过迭代之后，不仅可以建立更加精确的mask，也可以弥补特征信息的缺失。

The separation network of the proposed MSFFT-Net aims to exactly estimate a mask for each source sound from  $\mathbf{Z}$. Different from\cite{Luo2020DualPathRE}\cite{Chen2020DualPathTN}, our proposed separation network comprises $N$ stacked blocks, each of which is a multi-scale structure. As shown in Fig.\ref{fig:3}, each block consists of two parallel separation processing path, both of which contains two stacked speech separation transformers (SS-Transformer). In each block, the top separation processing path consists of intra- and inter chunk operation iteratively and alternatively, aiming to capture the low-level features with small scale while the bottom separation processing path focuses on capturing the high-level features with big scale. High-level feature and low-level feature are simultaneously processed in parallel through different processing paths. The intermediate outputs of both parallel paths are delivered to each other, enabling information from speech sequences at different scales could dynamically interact. We very in the following experiments that the information interaction mechanism can resolve the problem of feature resolution attenuation in each processing path. 

The decoder is finally used to reconstruct the separated speech signals in the time domain by using the masks predicted by the separation network. In brief, the $c$-th source sound signal can be reconstructed by applying the $c$-th estimated mask and the initial frame $\mathbf{\tilde{X}}$, then we simply use a 1-D transposed convolutional layer back to the waveform, which can be formulated as follows,
\begin{equation}
    \label{eq18}
   \mathbf{S}_c=ConvTrans1D(\mathbf{\tilde{X}}\odot \mathbf{M_c})
\end{equation}
where $\mathbf{M_c}$ and $\mathbf{S_c}$ denote the $c$-th estimated mask and the $c$-th estimated source speech, respectively.

\subsection{Separation Network}
\label{sec:transformer}
\begin{figure}[htb]
    \centering
    \includegraphics[scale=0.3]{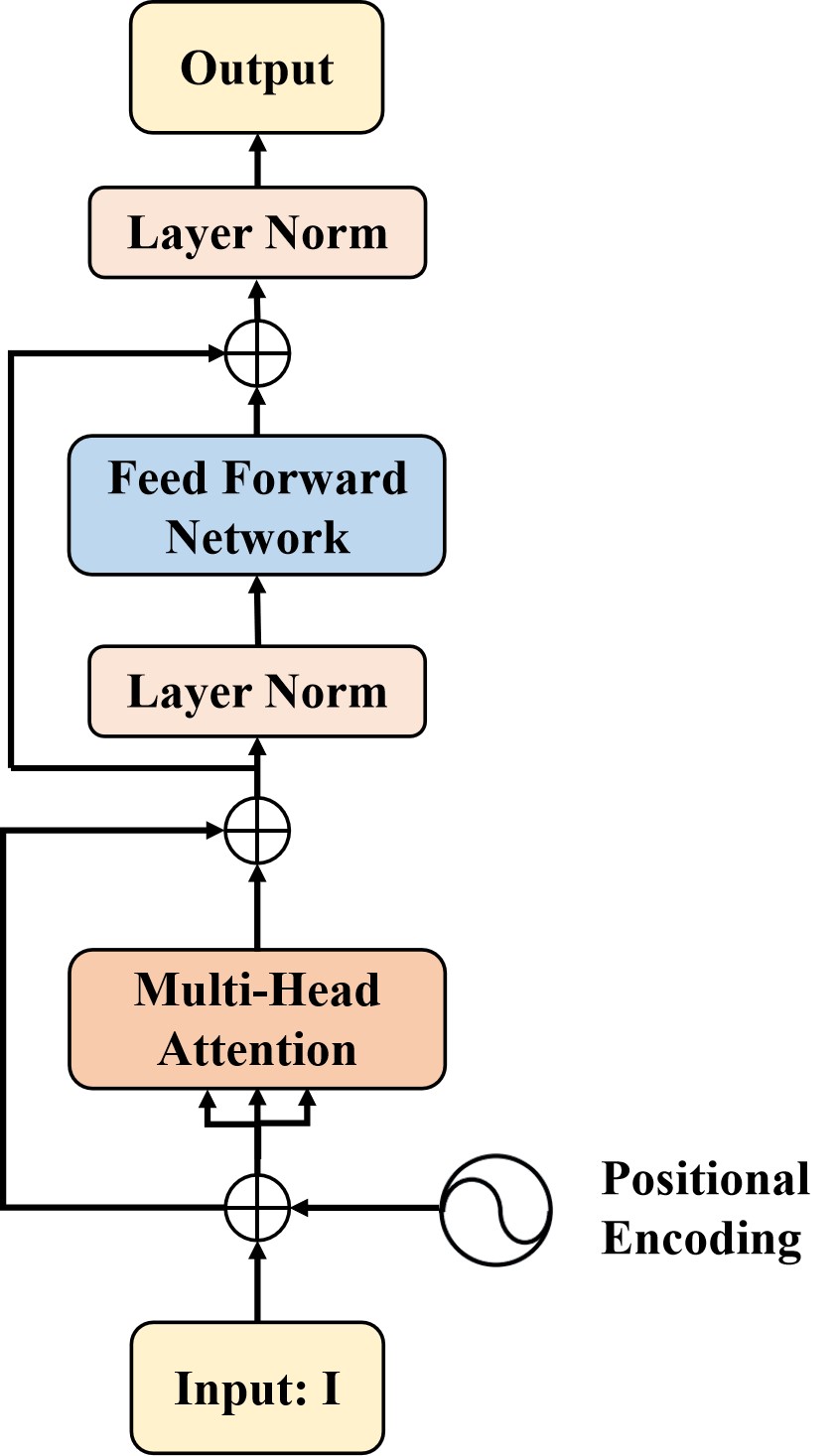}
    \caption{The overall architecture of original transformer encoder}
    \label{fig:4}
\end{figure}

In recent years, transformer has been widely applied for resolving the problem of limited receptive fields presented in CNN and RNN \cite{Vaswani2017AttentionIA, Yu2022SETransformerSE, Devlin2019BERTPO, Li2021PoseRW, Yuan2021HRFormerHT, Srinivas2021BottleneckTF, Lin2021CATCA}. The transformer typically consists of two parts: encoder and decoder. In this paper, speech separation transformer especially refers to the encoder part of transformer. Specifically, the speech separation transformer consists of $N_{intra}$ Intra-transformers and $N_{inter}$ Inter-transformers. Both Intra-transformers and Inter-transformers, as depicted in Fig.~\ref{fig:4}, are similar to the original transformer encoder part, which consists of two core components: multi-head attention (MHA) and position-wise feed-forward network (FFN).  

For the proposed MSFFT-Net, the detailed separation process can be divided into the following four stages.
\begin{itemize}
    \item Firstly, the 3-D feature tensor $\mathbf{Z}$ is passed to the first SS-Transformer of the first layer. Then, the output is down-sampled and passed to the first SS-Transformer of the second layer.
    \item Secondly, the feature tensor obtained from the first layer would be down-sampled and fused into the feature tensor obtained from the second layer; Meanwhile, the feature tensor obtained from the second layer would be up sampled and fused into the feature tensor obtained from the first layer. 
    \item Thirdly, these two 3-D feature tensor would be passed to the second SS-Transformer of the first layer and the second layer respectively. 
    \item Finally, the feature tensor obtained from the second layer would be up sampled and concatenated into the feature tensor obtained from the first layer and convert into 3-D tensor that is denoted by $\mathbf{Z}' \in \mathbb{R}^{D \times S \times K}$ via convolutional operation. 
\end{itemize}

Given an input $\mathbf{I}$, the overall architecture of original transformer encoder can be formulated as follows,
\begin{equation}
    \label{eq7}
    \mathbf{MHA}=MultiHeadAttention(\mathbf{I})
\end{equation}
\begin{equation}
    \label{eq8}
    \mathbf{LN}=LayerNorm(\mathbf{I} + \mathbf{MHA})
\end{equation}
\begin{equation}
    \label{eq9}
    \mathbf{FFN}=FeedForword(\mathbf{LN})
\end{equation}
\begin{equation}
    \label{eq10}
    \mathbf{Output}=LayerNorm(\mathbf{LN} + \mathbf{FFN})
\end{equation}

\begin{figure}[htpb]
    \centering
    \includegraphics[scale=0.3]{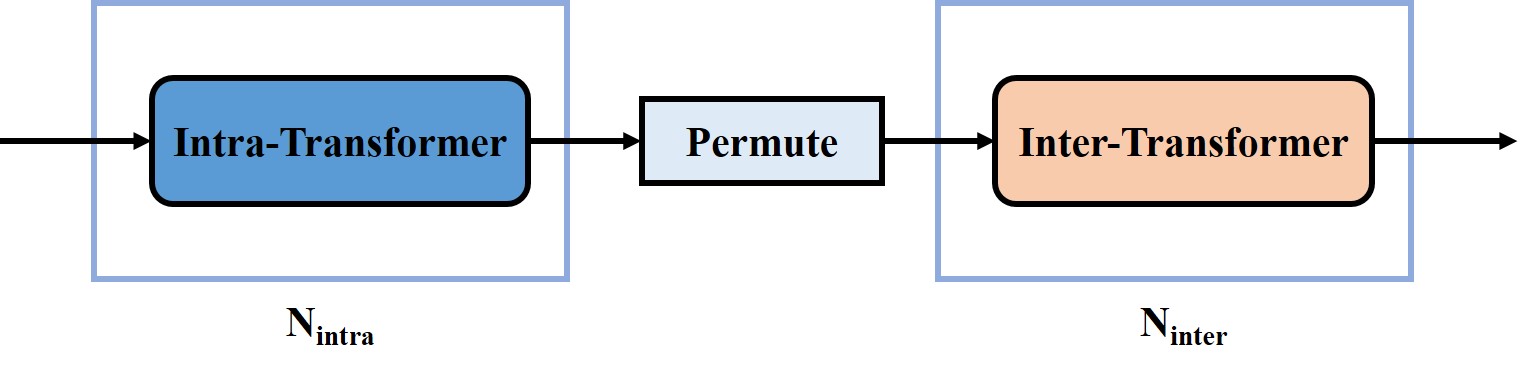}
    \caption{The overall architecture of speech separation transformer.}
    \label{fig:5}
\end{figure}

Fig.~\ref{fig:5} shows the overall architecture of the speech separation transformer of the proposed MSFFT-Net. $N_{intra}$ Intra-transformer is firstly applied to individual chunks in parallel to process local information, followed by permuting the output 3D-tensor, and then $N_{inter}$ Inter-transformer is applied across the chunks to capture global information. 

\begin{figure}[htb]
    \centering
    \begin{subfigure}[b]{0.3\textwidth}
        \includegraphics[scale=0.3]{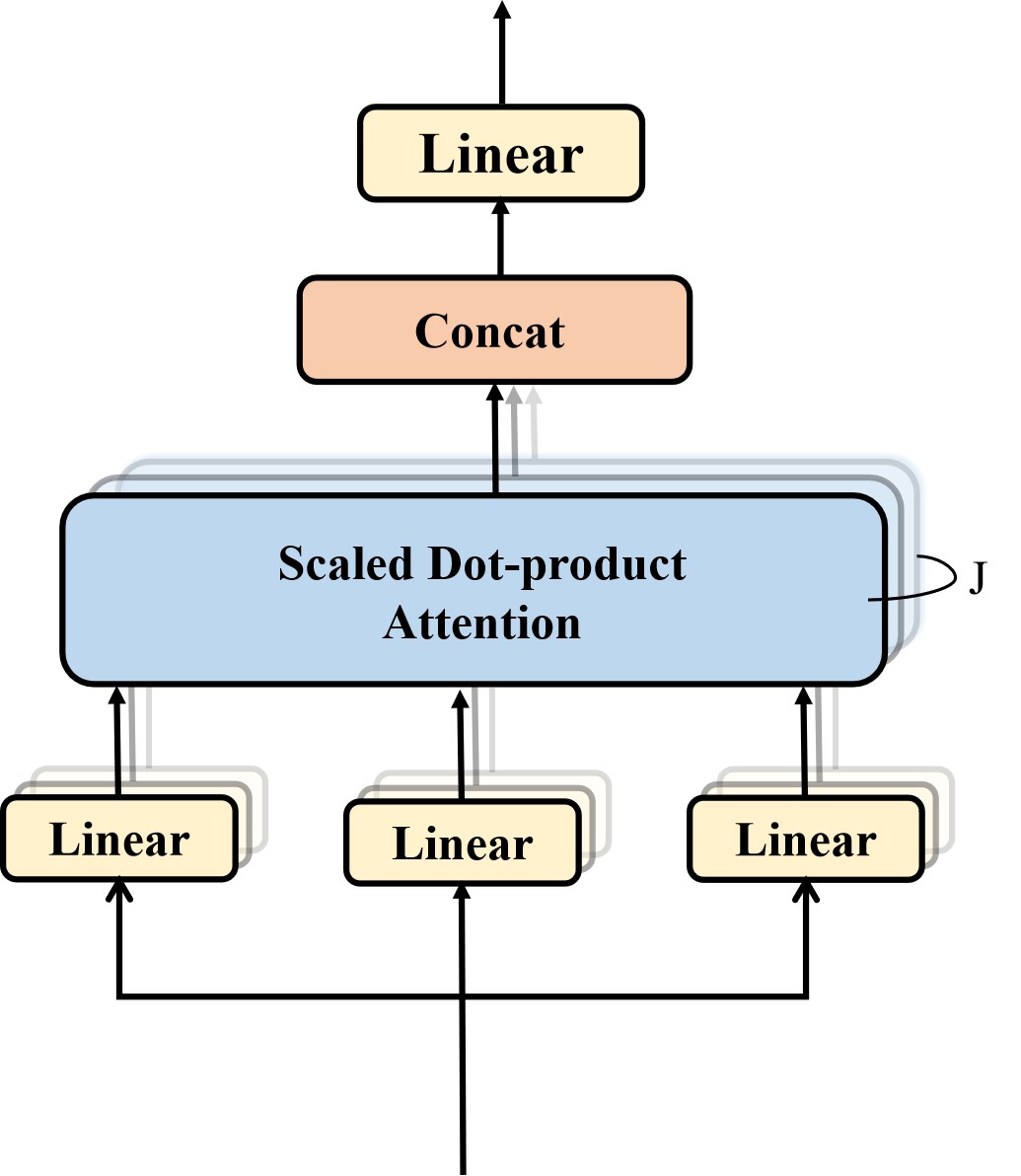}
        \caption{The processing of multi-head attention operation.}
        \label{fig:6a}
    \end{subfigure}
    \begin{subfigure}[b]{0.3\textwidth}
        \includegraphics[scale=0.3]{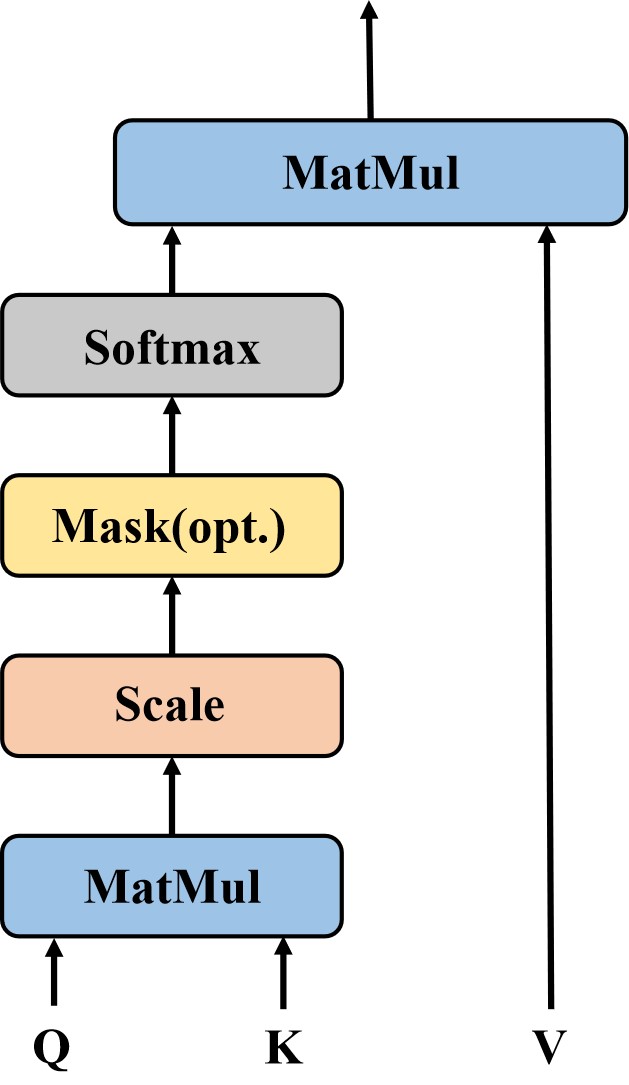}
        \caption{The calculation process of scaled dot-product attention}
        \label{fig:6b}
    \end{subfigure}
    
    \caption{Overall Architecture of attention mechanism of speech separation transformer.}
    \label{fig:6}
\end{figure}

Multi-head Attention has been proven to be an exactly effective self-attention mechanism to learn the long-term dependencies, which is shown in Fig.~\ref{fig:6} (a). More specifically, instead of performing a single attention function with values, keys and queries, multi-head attention attends to linearly project $value$, $key$ and $query$ $J$ times with different positions to refine the final representation. The processing of multi-head attention can be represented as,

\begin{equation}
    \label{eq11}
    head_j=A(Q_j,K_j,V_j),j\in [1,J]
\end{equation}
\begin{equation}
    \label{eq12}
    \mathbf{MultiHead}=Concat(head_1,...,head_J)\mathbf{W}^O
\end{equation}
where A($\cdot$) is the scaled dot-product attention operation as depicted in Fig.~\ref{fig:6} (b), and $W^O$ is a linear projection. Given $query$,$key$ and $value$ inputs of $J$ different heads, the scaled dot-product attention can be derived as,
\begin{equation}
    \label{eq13}
    A(Q_j,K_j,V_j)=Softmax(\frac{Q_j^T K_j}{\sqrt{D / J}})V_j,j\in [1,J]
\end{equation}
where $Softmax$($\cdot$) is adopt to convert the linear value into class probability.

\begin{figure*}[htpb]
    \centering
    \includegraphics[scale=0.5]{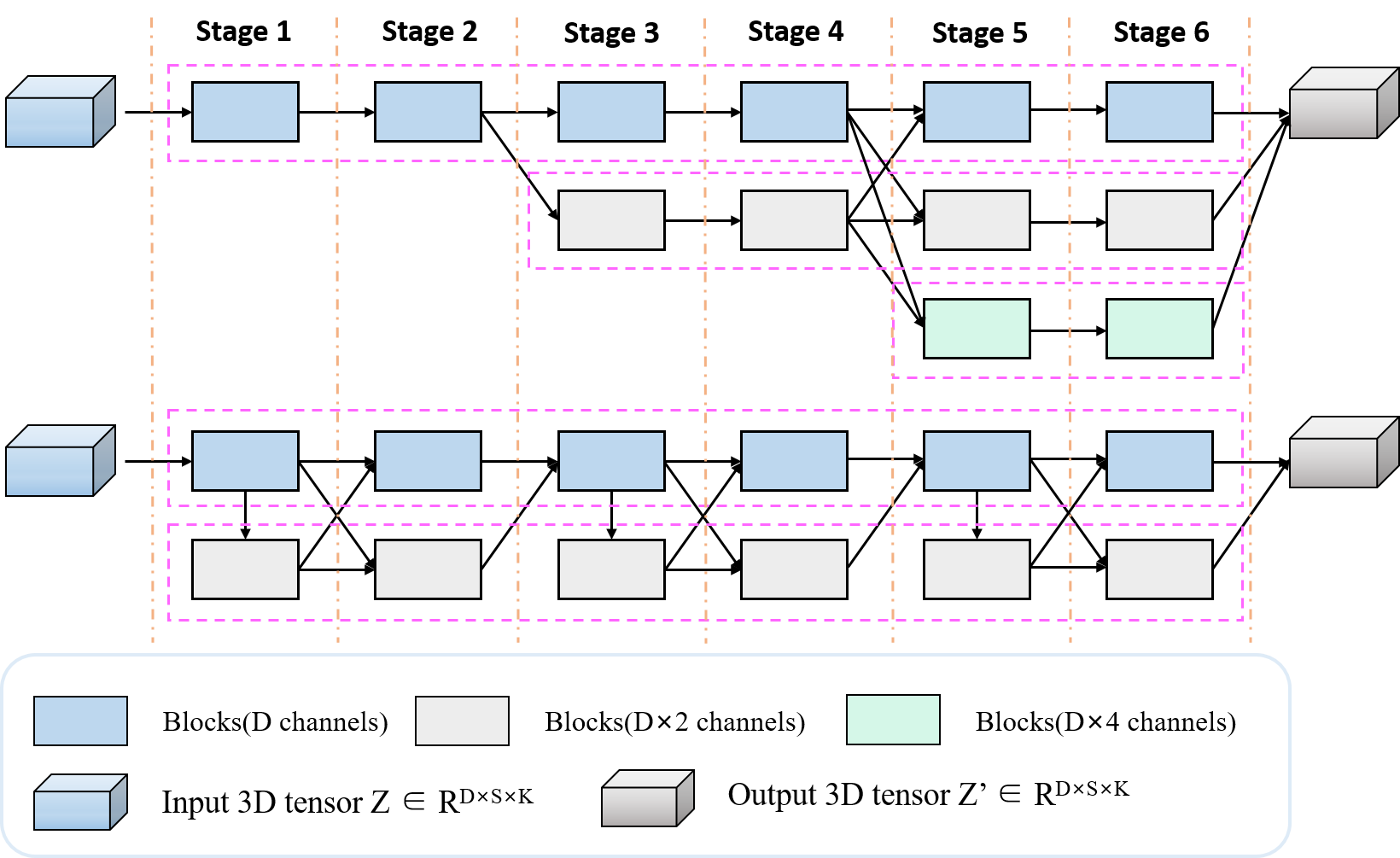}
    \caption{The structure of the proposed multi-scale feature fusion networks. (Top) The structure of MSFFT-3P. (Bottom) The structure of MSFFT-2P.}
    \label{fig:7}
\end{figure*}

\subsection{Multi-scale Feature Fusion}
\label{sec:MSFF}
Traditional dual-path based models such as DPRNN keep feature dimension and segment size unchanged throughout all layers of the network. Inspired by recent progress in the field of human posture estimation\cite{Newell2016StackedHN, Lu2021AFA} \cite{Sun2019DeepHR}, and multi-scale feature fusion approach \cite{Zhao2020MultiScaleGT} \cite{Thakur2022GradientAM, Liu2021AN2, Gao2021MultiscaleFB, Xu2021MonocularDE, Liu2021DenserNetWS}, we propose in this paper a novel multi-path feature fusion structure. Particularly, our proposed framework has multiple feature processing paths, each of which corresponds to feature processing with different feature scales. The feature obtained from the upper path would be down-sampled and fused into the feature map of the lower path. The feature obtained from the lower path would also be up-sampled and fused into the feature map of the upper path. The final output of each path are fused together to form one feature map.

In this paper, two types of multi-scale feature fusion structures as shown in Fig.~\ref{fig:7} (a) and Fig.~\ref{fig:7} (b) are proposed. Specifically, the top pannel of Fig.~\ref{fig:7} plots the first type of the proposed multi-scale feature fusion structure (denoted as MSFFT-3P) which contains three paths. The second type of multi-scale feature fusion structure (denoted as MSFFT-2P) is plot in the bottom pannel of Fig.~\ref{fig:7}, where only two paths are configured to perform feature fusion. 

Both of the two feature fusion structures consist of down-sampling, up-sampling and fusion operations. Information of adjacent paths can be inter-exchanged with each other. Down-sampling and up-sampling operations can be formulated as,
\begin{equation}
    \label{eq14}
    DS(\mathcal{X})=Conv1D_K(\mathcal{X})
\end{equation}
\begin{equation}
    \label{eq15}
    US(\mathcal{X})=ConvTrans1D_K(\mathcal{X})
\end{equation}
where $\mathcal{X}$ denotes a tensor $\in \mathbb{R}^{D \times S \times K}$, $DS$($\cdot$) and $US$($\cdot$) denote down-sampling and up-sampling operations, respectively, $Conv1D_K$($\cdot$) and $ConvTrans1D_K$($\cdot$) denote 1-D convolution and 1-D transposed convolution operations along the axis of $K$ respectively. In this paper, in order to improve the computation efficiency, element-wise sum operation is adopt as the fusion operation instead of concatenation operation. In the MSFF-3P structure, the Information exchanging occurred between adjacent paths can be formulated as,
\begin{equation}
    \label{eq16}
    \mathcal{X}_{s+1}^{p}=\begin{cases} 
    \mathcal{Y}_s^{p}\oplus US(\mathcal{Y}_s^{p+1}), p=1\\ 
    \mathcal{Y}_s^{p}\oplus DS(\mathcal{Y}_s^{p-1}), p=2\\ DS(\mathcal{Y}_s^{p-2})\oplus DS(\mathcal{Y}_s^{p-1}), p=3
    \end{cases}  
\end{equation}
where $\oplus$ denotes element-wise sum operation, $\mathcal{Y}_s^{p}$ denotes the output tensor for the $s$-th stage of the $p$-th path, and $\mathcal{X}_{s+1}^{p}$ denotes the input tensor for the $(s+1)$-th stage of the $p$-th path.

Notice that the number of stages in MSFFT-2P should be even, and information exchange takes place after odd stages. Instead of employing element-wise sum operation as the fusion operation throughout all stages of the network in MSFFT-3P, we employ concatenating operation as fusion operation after even stages in MSFFT-2P. Compared with element-wise sum operation, concatenating operation can significantly reduce the loss of feature information. The fusion processing in the MSFFT-2P structure can be formulated as,
\begin{equation}
    \label{eq17}
    \mathcal{X}_{s+1}^{p}=\begin{cases} 
    \mathcal{Y}_s^{p}\oplus US(\mathcal{Y}_s^{p+1}), p=1, s\%2=1\\ 
    \mathcal{Y}_s^{p}\oplus DS(\mathcal{Y}_s^{p-1}), p=2, s\%2=1\\ 
    Concat(\mathcal{Y}_s^{p}, \mathcal{Y}_s^{p+1}), p=1, s\%2=0
    \end{cases}  
\end{equation}

\subsection{Utterance-level based Permutation Invariant Training}
\label{sec:uPIT}
Speech separation task has multiple outputs for a given source. The training set consists of a lot of utterances spoken by multiple speakers. In this paper, we permute speaker labels to the corresponding outputs with Permutation invariant training (PIT)\cite{Yu2017PermutationIT} where the mean square error (MSE) between the estimated waveform and its corresponding counterpart is adopt as the cost function, which can be formulated as follows,
\begin{equation}
\begin{split}
\label{eq18}
J_x&=\frac{1}{T\times F \times C}\sum_{c=1}^{C}\left \| \hat{S_c}-S_c \right \|^2  \\
&=\frac{1}{T\times F \times C}\sum_{c=1}^{C}\left \| M_c\circ \tilde{X} - S_c\right \|^2
\end{split} 
\end{equation}
where $T$ and $F$ denote the number of time frames and frequency bins respectively, and $\circ$ is a binary operation, in this paper, we refer $\circ$ to element-wise multiplication.

In order to effectively solve the speaker tracing problem \cite{Kolbaek2017MultitalkerSS}, in this paper, we adopt an utterance-level cost function to the conventional PIT which can be rewritten as follows,
\begin{equation}
\begin{split}
\label{eq19}
J_{\phi ^*} &=\frac{1}{V}\sum_{c=1}^{C}\left \| M_c\circ \tilde {X} -S_{\phi ^*(c)} \circ cos(\theta _y-\theta _{\phi ^*(c)}) \right \|_F^2
\end{split} 
\end{equation}
where $V=T\times F \times C$, $\phi^*$ denotes the permutation set of $C$ speakers and $\| \cdot \|_F$ denotes the Frobenius norm.

\section{Experiments}
\label{sec:experiments}

\subsection{Datasets}
\label{sec:data}
In order to verify the effect of the proposed mutli-scale feature fusion method, the MSFF-Net was used to perform speech separation on two benchmark datasets, i.e., WSJ0-2mix and WSJ0-3mix, both of which are generated from the Wall Street Journal (WSJ0) corpus \cite{Hershey2016DeepCD} by randomly selecting clean utterances from different speakers and mixing them at random signal-to-noise ratios (SNR) ranges from -5dB to 5dB. The WSJ0 consists of 30 hours training data, 10 hours validation data and 5 hours test data.

\subsection{Evaluation Metrics}
\label{sec:evaluation}
We employ two popular indicators to evaluate the the separated speech in terms of signal fidelity as well as audibility, that is, the the scale-invariant source-to-noise ratio improvement (SI-SNRi) indicator and the signal-to-distortion ratio improvement (SDRi) indicator. The SI-SNR is defined as follows,
\begin{equation}
    \label{eq20}
   SI-SNR=10log_{10}\frac{\|s^{target}\|^2}{\|s^{noise}\|^2}
\end{equation}
where $s^{target}=\Pi_{s}{(\hat{s})}$, $s^{noise}=\hat{s}-s^{target}$, $\Pi_{s}{(\hat{s})}$ is the projection of $\hat{s}$ on $s$, $s$ and $\hat{s}$ are the reference target speech and the estimated speech, respectively. 

Given a mixed speech $x$, the SDRi and the SI-SNRi can be defined as:
\begin{equation}
    \label{eq21}
    SDRi(\hat{s},s,x)=SDR(\hat{s},s)-SDR(x,s)\\
\end{equation}
\begin{equation}
    \label{eq21}
    SI-SNRi(\hat{s},s,x)=SI-SNR(\hat{s},s)-SI-SNR(x,s)
\end{equation}

\subsection{Training details}
In order to confirm the effectiveness of the proposed MSFFT-Net method, two types of MSFFT-Net, i.e., the MSFFT-3P and the MSFFT-2P are implemented to perform speech separation. Existing state-of-the-art separation methods such as the dual-path RNN \cite{Luo2020DualPathRE} (denoted as DPRNN), the dual-path Transformer Network \cite{Chen2020DualPathTN} (denoted as DPTNet) as well as the Sepformer proposed in \cite{Subakan2021AttentionIA} are also used for speech separation performance comparison. All methods are trained on NVIDIA V100 GPU with 32GB of memory. 
\begin{table}[htb]
    \centering
    \caption{The parameter definitions of different separation methods.}
    \label{table1}
    \begin{tabular}{@{}c|c@{}}
    \hline
    Symbol  &  Description \\
    \hline \hline
    $E$  & Number of filters in autoencoder\\
    \hline
    $D$  & Number of filters\\
    \hline
    $N_{intra}$  & Number of intra-transformer layers in each block\\
    \hline
    $N_{inter}$  & Number of inter-transformer layers in each block\\
    \hline
    $N$ & Number of training stages\\
    \hline
    $M$ & The window length\\
    \hline
    $J$ & Number of attention heads\\
    \hline
    $K$ & The segment size\\
    \hline
    $DFF$ & Dimension of positional feed-forward network\\
    \hline
    \end{tabular}
\end{table}
The parameter definitions of different separation methods are described in Table.\ref{table1} and the overall parameter configuration details are shown in Table.\ref{table2}.
\begin{table*}[htb]
  \centering
  \caption{Parameter configurations of different speech separtion methods.}
    \label{table2}
    \begin{tabular}{c|cccccccccc}
    \hline
    Model & E & D & {${N}_{intra}$} & ${N_{inter}}$ &N & M & Stride & J & K & DFF  \\
    \hline
    DPRNN\cite{Luo2020DualPathRE} & 256   & 64    & 1     & 1     & 6     & 2     & 1     & -     & 250   & -     \\
    \hline
    DPTNet\cite{Chen2020DualPathTN} & 256   & 64    & 1     & 1     & 6     & 2     & 1     & 4    & 250   & -     \\
    \hline
    Sepformer\cite{Subakan2021AttentionIA} & 256   & 256   & 4     & 4     & 2     & 16    & 8     & 8     & 250   & 2048   \\
    \hline
    MSFFT-3P(ours)  & 512   & 128   & 2     & 2     & 6     & 16    & 8     & 16    & 256   & 2048  \\
    \hline
    MSFFT-2P(ours)  & 512   & 128   & 4     & 4     & 2     & 16    & 8     & 16    & 256   & 2048 \\
    \hline
    \end{tabular}%
  
\end{table*}%

\subsubsection{The proposed MSFFT-Net}
For the proposed MSFFT-Net, we set $E=512$, $D=128$ for MSFFT-3P and MSFFT-2P respectivelly. In each convolutional filter, the window length $M$ is set to 16 sample points and the stride is set to 8. As for separation network, the initial segment size $K$ is set to 256, and the total stages $N$ is set to 6 for MSFFT-3P and 2 for MSFFT-2P, respectively. For each block, we employ $N_{intra}$ Intra-transformers and $N_{inter}$ Inter-transformers where both $N_{intra}$ and $N_{inter}$ are equal to 4 for MSFFT-3P and 2 for MSFFT-2P. For the MSFFT-3P, the $s$ in eq.(~\ref{eq16}) is set to 4 as information exchange only occurs after the $4$-th stage in MSFFT-3P structure. For the MSFFT-2P, the $s$ in eq.(~\ref{eq17}) is set to 2,4,6 in the following experiments. In each SS-Transformer, 16 parallel attention heads and 2048-dimension positional feed-forward networks are used, i.e. $J=16$ and $DFF=2048$. Adam algorithm \cite{Subakan2021AttentionIA} is used as the optimizer and gradient clipping with a maximum L2-norm of 5. The initial learning rate is set to $4e^{-4}$. Both of the MSFFT-3P and MSFFT-2P models are trained with 200 epochs in total. After 85th epochs, we halve the learning rate if no performance improvement is observed in the validation set for 3 consecutive epochs. 

\subsubsection{DPRNN}
A dual-path RNN \cite{Luo2020DualPathRE} consists of segmentation, block processing and overlap-add. The segmentation stage splits a sequential input into overlapped chunks and concatenates all the chunks into a 3-D tensor. Then the tensor is passed to the stacked DPRNN blocks. Each block contains two sub-blocks corresponding to intra-block and inter-block processing respectively. specifically, both the intra-block and inter-block processing employ Bi-LSTM and are performed iteratively and alternatively within the DPRNN block. The output of the last layer is transformed to a sequential output using the overlap-add method. 

%按照\subsubsection{The proposed MSFFT-Net}节的描述方法将DPRNN的配置用文字描述。
In this paper, the configurations of DPRNN are as follows. In encoder-decoder modules, the number of filters in autoencoder is set to 256 and the number of filters is set to 64, and 6 consecutive DPRNN blocks are used in the separation network for masking estimated. The window length and segment size are set to 2 and 250, respectively.

\subsubsection{DPTNet}
Like DPRNN, the DPTNet\cite{Chen2020DualPathTN} also consists of  encoder, dual-path transformer processing and decoder. In the dual-path transformer processing stage, DPTNet employ a modified transformer-based structure instead of employing Bi-LSTM, which removes the positional encoding layer and replaces the first linear layer within the feed-forward network in the original transformer with an RNN structure to capture the positional information of speech sequences. These modifications leads to better results than the DPRNN, but also leads to additional computational cost much heavier than DPRNN, which make it difficult to apply in realistic scenarios.

%按照\subsubsection{The proposed MSFFT-Net}节的描述方法将DPTNet的配置用文字描述。
In this paper, the configurations of DPTNet is the same as the that of the DPRNN except that 4 parallel attention heads are used in each modified transformer in the DPTNet.

\subsubsection{Sepformer}
The Sepformer proposed in \cite{Subakan2021AttentionIA} contains several blocks to process intra-chunk and inter-chunk respectively. Each block applies $N_{intra}$ Intra-Transformer layers to model short-term dependencies within each chunk and $N_{inter}$ Inter-Transformer layers to model long-term dependencies across the chunks. 

%按照\subsubsection{The proposed MSFFT-Net}节的描述方法将Sepformer的配置用文字描述。
In this paper, the configuration of the Sepformer are as follows. In encoder-decoder modules, the number of filters in autoencoder and the number of filters are both set to be 256, and 6 consecutive Sepformer blocks are used in separation network. The window length and segment size are set to be 16 and 250 respectively. In each intra-transformer and inter-transformer, 8 paralell attention heads and 2048-dimension postional feed-forward networks are employed.
\section{Results}
\label{sec:results}

\subsection{Comparison with the state-of-the-art speech separation methods}
\label{sec:comparison}

Table~\ref{table:3} reports the SI-SNRi and SDRi scores of the proposed MSFFT-NET and other baselines on the WSJ0-2mix dataset. In table \ref{table:3}, the methods proposed in \cite{Luo2018TaSNetTA}\cite{Luo2019ConvTasNetSI}\cite{FurcaNeXt} \cite{Luo2020DualPathRE}\cite{Nachmani2020VoiceSW}\cite{Chen2020DualPathTN}\cite{Subakan2021AttentionIA}\cite{Lam2021EffectiveLT}\cite{Lam2021SandglassetAL} belong to the type of Tasnet method and the remainders belong to the type of the T-F domain based method. As can be seen in Table~\ref{table:3}, the proposed MSFFT-3P and MSFFT-2P achieve 20.7 dB and 21 dB in SI-SNRi, respectively, both obtaining better performance and high efficiency compared with other popular Tasnets. Particularly, the proposed MSFFT-2P obtains the new state-of-the-art speech separation performance.
 
\begin{table}[htpb]
    \centering
    \caption{Speech separation performance of different methods on WSJ0-2mix dataset.}
    \label{table:3}
    {
    \centering
    \begin{tabular}{@{}cccc@{}}
    \toprule
    Model  &  SI-SNRi(dB)  & SDRi(dB) &  Model Size(M) \\ 
    \midrule
    DPCL++,2016\cite{Isik2016SingleChannelMS} & 10.8  &  11.2 &  13.6 \\
    uPIT-BLSTM-ST,2017\cite{Kolbaek2017MultitalkerSS} & 9.8  &  10.0 &  92.7 \\
    Chimera++,2018\cite{Wang2018AlternativeOF} & 11.5  &  12.0 &  32.9 \\
    BLSTM-TasNet,2018\cite{Luo2018TaSNetTA} & 13.2  &  13.6 &  23.6 \\
    Conv-TasNet,2019\cite{Luo2019ConvTasNetSI} & 15.3  &  15.6 &  5.1 \\
    Deep CASA,2019\cite{Liu2019DivideAC} & 17.7  &  18.0 &  12.8 \\
    FurcaNeXt,2019\cite{FurcaNeXt} & -  &  18.4 &  51.4 \\
    DPRNN,2020\cite{Luo2020DualPathRE} & 18.8  &  19.0 &  2.6 \\
    Gated DPRNN,2020\cite{Nachmani2020VoiceSW} & 20.1  &  - &  7.5 \\
    DPTNet,2020\cite{Chen2020DualPathTN} & 20.2  &  20.6 &  2.7 \\
    Sepformer,2021\cite{Subakan2021AttentionIA} & 20.4  &  20.5 &  26 \\
    GALR,2021\cite{Lam2021EffectiveLT} & 20.3  &  20.5 &  \textbf{2.3} \\
    Sandglasset(MG),2021\cite{Lam2021SandglassetAL} & 20.8  &  21.0 &  \textbf{2.3} \\
    MSFFT-3P(ours)  &  20.7  &  20.8 & 34.4 \\
    %\textbf{MSFFT-2P}  &  20.8  &  21.0 & 30.9 \\
    MSFFT-2P(ours)  &  \textbf{21.0}  &  \textbf{21.2} & 30.9 \\
    \toprule
    \end{tabular}
    }
\end{table}

\begin{figure*}[htb]
    \centering
    \begin{subfigure}[b]{0.22\textwidth}
        \includegraphics[scale=0.6]{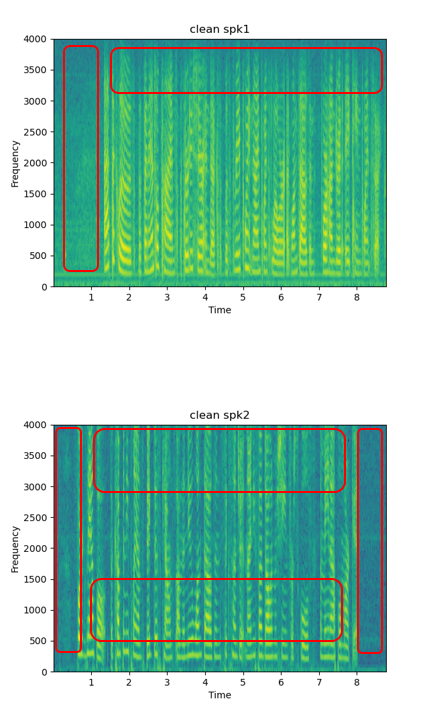}
        \caption{}
        \label{fig:8a}
    \end{subfigure}
    \begin{subfigure}[b]{0.22\textwidth}
        \includegraphics[scale=0.6]{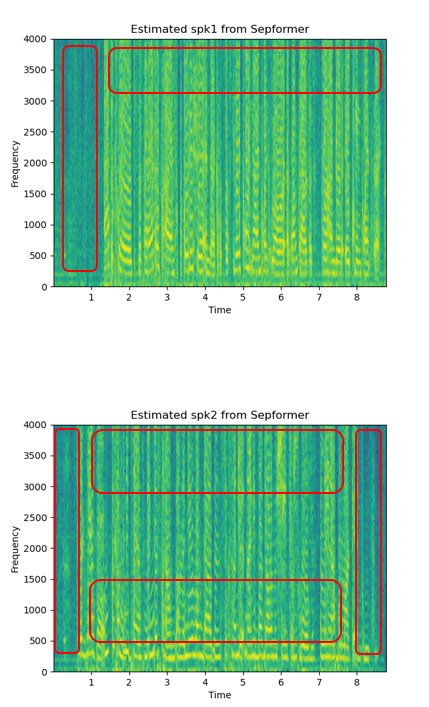}
        \caption{}
        \label{fig:8b}
    \end{subfigure}
    \begin{subfigure}[b]{0.22\textwidth}
        \includegraphics[scale=0.6]{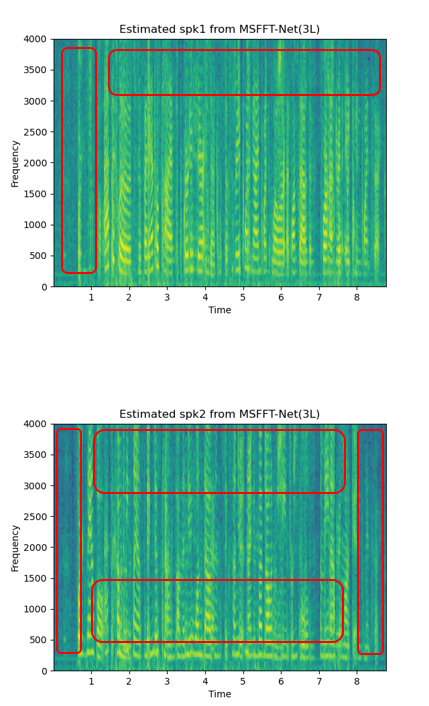}
        \caption{}
        \label{fig:8c}
    \end{subfigure}
    \begin{subfigure}[b]{0.22\textwidth}
        \includegraphics[scale=0.6]{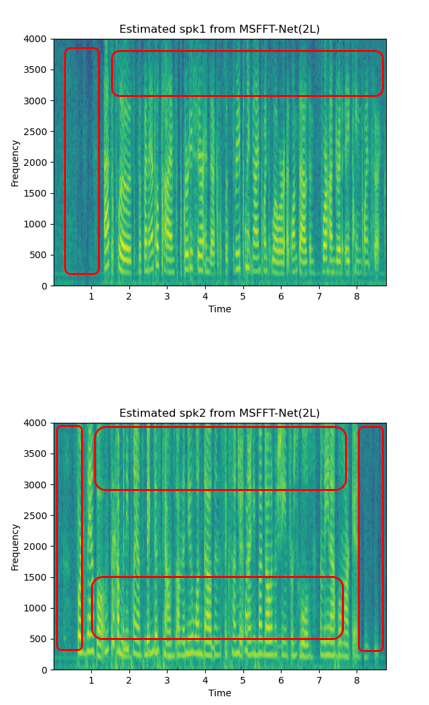}
        \caption{}
        \label{fig:8d}
    \end{subfigure}
    \caption{The spectrograms estimated by different speech separation methods. (a) clean speech spectrograms of 1\# (top) and 2\#(bottom) speakers in the mixed speech. (b) Spectrograms of 1\  (top) and 2\# (bottom) speakers estimated by the Sepformer method. (c) Spectrogram of 1\# (top) and 2\# (bottom) speakers estimated by the MSFFT-3P. (d) Spectrogram of 1\# (top) and 2\# (bottom) speakers estimated by the MSFFT-2P.}
    \label{fig:8}
\end{figure*}

For subjective evaluation, We give a speech separation example in Fig.\ref{fig:8}, where the speech spectrograms separated by the proposed MSFFT-Net and Sepformer \cite{Subakan2021AttentionIA} are plot respectively. From Fig.~\ref{fig:8}, we can see that the two speaker speech spectrograms separated by MSFFT-3P and MSFFT-2P are both closer to that of the clean original speech than that obtained by the Sepformer method, indicating effectiveness of feature fusion on speech separation performance improvements of the proposed MSFFT-Net.

In addition, we evaluate in depth the performance of the proposed MSFFT-Net on the WSJ0-3mix dataset, where each mixing speech comes from the utterances of three different speakers. Since MSFFT-2P obtains a better performance than MSFFT-3P on WSJ0-2mix, we just train MSFFT-2P on WSJ0-3mix. As shown in Table~\ref{table:4},  MSFFT-2P obtains a SI-SNRi of 18.1dB and a SDRi of 18.4dB, which again achieve new state-of-the-arts without any data augmentation on WSJ0-3mix dataset, indicating the effectiveness and generality of the proposed MSFFT-Net. 

\begin{table}[htb]
    \centering
    \caption{Comparison with other methods on WSJ0-3mix}
    \label{table:4}
    {
    \centering
    \begin{tabular}{@{}cccc@{}}
    \toprule
    Model  &  SI-SNRi(dB)  & SDRi(dB) &  Model Size(M) \\ 
    \midrule
    Conv-TasNet,2019\cite{Luo2019ConvTasNetSI} & 12.7  &  13.1 &  5.1 \\
    DPRNN,2020\cite{Luo2020DualPathRE} & 14.7  &  - &  2.6 \\
    Gated DPRNN+Spk ID,2020\cite{Nachmani2020VoiceSW} & 16.7  &  - &  7.5 \\
    Wavesplit+Spk ID,2021\cite{Zeghidour2021WavesplitES} & 17.3  &  17.6 &  42.5 \\
    Sepformer,2021\cite{Subakan2021AttentionIA} & 17.6  & 17.9 &  26 \\
    Sandglasset(MG),2021\cite{Lam2021SandglassetAL} & 17.1  &  17.4  &  \textbf{2.3} \\
    MSFFT-2P(ours)  &  \textbf{18.1}  &  \textbf{18.4} & 30.9 \\
    \toprule
    \end{tabular}
    }
\end{table}

\subsection{Ablation study}
\label{sec:ablation}
In this section, we evaluate the performance of the proposed MSFFT-NET with different parameter configurations and different fusion strategies. Table.\ref{table:22} gives the speech separation performance of the MSFFT-2P and the MSFFT-3P with different parameter configurations. One can seen that the MSFFT-2P outperforms the MSFFT-3P. This may be attribute to the larger value setting of ${N}_{intra}$ and ${N}_{inter}$ in the MSFFT-2P.

\begin{table*}[htbp]
  \centering
  \caption{Performance evaluation of MSFFT-Net on WSJ0-2Mix with respect to different parameter configurations}
  \label{table:22}%
    \begin{tabular}{c|cccccccccc|c}
    \hline
    \textbf{Model} & \textbf{E} & \textbf{D} & $\textbf{N}_{intra}$ & $\textbf{N}_{inter}$ & \textbf{N} & \textbf{M} & \textbf{Stride} & \textbf{J} & \textbf{K} & \textbf{DFF} & \textbf{SI-SNRi(dB)} \\
   \hline
    \multirow{4}[2]{*}{MSFFT-3P} & 256   & 64    & 1     & 1     & 6     & 16    & 8     & 4     & 256   & 2048  & 18.2 \\
          & 256   & 64    & 1     & 1     & 6     & 16    & 8     & 8     & 256   & 2048  & 18.5 \\
          & 512   & 128   & 2     & 2     & 6     & 16    & 8     & 8     & 256   & 2048  & 20.6 \\
          & 512   & 128   & 2     & 2     & 6     & 16    & 8     & 16    & 256   & 2048  & \textbf{20.7} \\
    \hline
    \multirow{4}[2]{*}{MSFFT-2P} & 256   & 64    & 1     & 1     & 6     & 16    & 8     & 4     & 256   & 2048  & 18.5 \\
          & 512   & 128   & 2     & 2     & 4     & 16    & 8     & 8     & 256   & 2048  & 20.6 \\
          & 512   & 128   & 4     & 4     & 2     & 16    & 8     & 8     & 256   & 2048  & 20.8 \\
          & 512   & 128   & 4     & 4     & 2     & 16    & 8     & 16    & 256   & 2048  & \textbf{21.0} \\
    \hline
    \end{tabular}%
\end{table*}%
For simplicity, we focus on the MSFFT-2P for fusion strategies study. In the MSFFT-2P, concatenation operation is adopt as the fusion operation after even stages. In this section, we replace fusion operation in all stages of MSFFT-2P with element-wise sum operation, and denote it as MSFFT-2P-S. The results reported in Table~\ref{table:5} show that MSFFT-2P obtain a better result than MSFFT-2P-S, which denotes the concatenation fusion method is better than the element-wise sum fusion method.

\begin{table}[htb]
    \centering
    \caption{The performance of the MSFFT-2P with respect to different fusion strategies on WSJ0-2Mix}
    \label{table:5}
    \begin{tabular}{@{}c|c|c|c|c|c|c@{}}
    \hline
   \textbf{Model}  & \textbf{D} & $\textbf{N}_{intra}$ & $\textbf{N}_{inter}$  &  \textbf{N}  & \textbf{J}  & \textbf{SI-SNRi} \\
    \hline \hline
    MSFFT-2P-S  &  64  &  1 & 1 & 6 & 4 & 18.2\\
    \hline
    MSFFT-2P  &  64  &  1 & 1 & 6 & 4 & \textbf{18.5}\\
    \hline
    MSFFT-2P-S  &  128  &  4 & 4 & 2 & 16 & 20.8\\
    \hline
    MSFFT-2P  &  128  &  4 & 4 & 2 & 16 & \textbf{21.0}\\
    \hline
    \end{tabular}
\end{table}

We also test the effectiveness of the proposed MSFFT-3P and MSFFT-2P with comparison to the Sepformer method. Table.~\ref{table:6} proves again the effectiveness of MSFFT-Net.

\begin{table}[htb]
    \centering
    \caption{The performance comparison with Sepformer on WSJ0-2Mix dataset.}
    \label{table:6}
    \begin{tabular}{@{}c|c|c|c|c|c|c@{}}
    \hline
    Model  & D & $N_{intra}$  &  $N_{inter}$  &  N  & J  & SI-SNRi \\
    \hline \hline
    Sepformer  &  64  &  1 & 1 & 6 & 4 & 17.4\\
    \hline
    MSFFT-3P  &  64  &  1 & 1 & 6 & 4 & \textbf{18.2}\\
    \hline
    Sepformer  &  128  &  4 & 4 & 2 & 16 & 20.3\\
    \hline
    MSFFT-2P  &  128  &  4 & 4 & 2 & 16 & \textbf{21.0}\\
    \hline
    \end{tabular}
\end{table}

In addition, we test the effect of the processing depth in MSFFT-2P, As seen in Table.~\ref{table:7}, there is a slight performance improvement when the depth is increased from 4 to 6. 

\begin{table}[htb]
    \centering
    \caption{The performance of MSFFT-Net on WSJ0-2Mix dataset with respect to different parameter configuration.}
    \label{table:7}
    \begin{tabular}{@{}c|c|c|c|c|c|c@{}}
    \hline
    Model  & D & $N_{intra}$  &  $N_{inter}$  &  N  & J  & SI-SNRi \\
    \hline \hline
    MSFFT-2P  &  128  &  2 & 2 & 4 & 8 & 20.7\\
    \hline
    MSFFT-2P  &  128  &  2 & 2 & 6 & 8 & \textbf{20.8}\\
    \hline
    \end{tabular}
\end{table}

\subsection{Embedding the proposed MSFF structure to DPRNN and DPTNet approaches}
\label{sec:transfer}
Our multi-scale feature fusion strategy is based on dual-path approach. Therefore, we try transplant multi-scale feature fusion structure to the DPRNN and DPTNet. As shown in Table~\ref{table:8}, we firstly apply the MSFF-3P structure to DPRNN and DPTNet respectively. The experimental results show improvements both in DPRNN and DPTNet, where MSFF-3P  RNN achieve 18.7dB (0.6dB higher than DPRNN) on the metric of SI-SNRi and MSFF-3P DPTNet achieve 19.4dB (0.4dB higher than that of the DPTNet).

\begin{table}[h]
    \centering
    \caption{The performance of DPRNN and DPTNet with MSFF-3P embedding on WSJ0-2Mix dataset.}
    \label{table:8}
    \begin{tabular}{@{}c|c|c|c|c|c@{}}
    \hline
    Model  &  E & D & M & K & SI-SNRi \\
    \hline \hline
    DPRNN  & 256 & 64 & 4 & 256 & 18.1 \\
    \hline
    MSFF-3P DPRNN  & 256 & 64 & 4 & 256 & \textbf{18.7} \\
    \hline
    DPTNet  & 256 & 64 & 4 & 256 & 19.0 \\
    \hline
    MSFF-3P DPTNet  & 256 & 64 & 4 & 256 & \textbf{19.4} \\
    \hline
    \end{tabular}
\end{table}

\begin{table}[h]
    \centering
    \caption{The performance of DPRNN and DPTNet with MSFF-2P embedding on WSJ0-2Mix dataset.}
    \label{table:9}
    \begin{tabular}{@{}c|c|c|c|c|c@{}}
    \hline
    Model  &  E & D & M & K & SI-SNRi \\
    \hline \hline
    DPRNN  & 256 & 64 & 8 & 128 & 16.8 \\
    \hline
    MSFF-2P DPRNN  & 256 & 64 & 8 & 128 & \textbf{17.9} \\
    \hline
    DPTNet  & 256 & 64 & 16 & 128 & 16.2 \\
    \hline
    MSFF-2P DPTNet  & 256 & 64 & 16 & 128 & \textbf{17.6} \\
    \hline
    \end{tabular}
\end{table}

We apply MSFF-2P structure to DPRNN and DPTNet respectively. From Table~\ref{table:9}, we also see significant improvements both in DPRNN and DPTNet, where MSFF-2P RNN achieve 17.9dB (1.1dB higher than DPRNN) on the metric of SI-SNRi and MSFF-2P DPTNet achieve 17.6dB (1.4dB higher than DPTNet). 

In summary, both MSFF-3P and MSFF-2P based on DPRNN and DPTNet obtain a better result than original DPRNN and DPTNet. These experimental results argue that our proposed multi-scale feature fusion methods are particularly effective. The features captured in different scales enrich more speaker information, making it obtain great speech separation performance improvement.

\subsection{Complexity analysis of different speech separation methods}
\label{sec:comparison}

\begin{figure}[htb]
    \centering
    \includegraphics[scale=0.5]{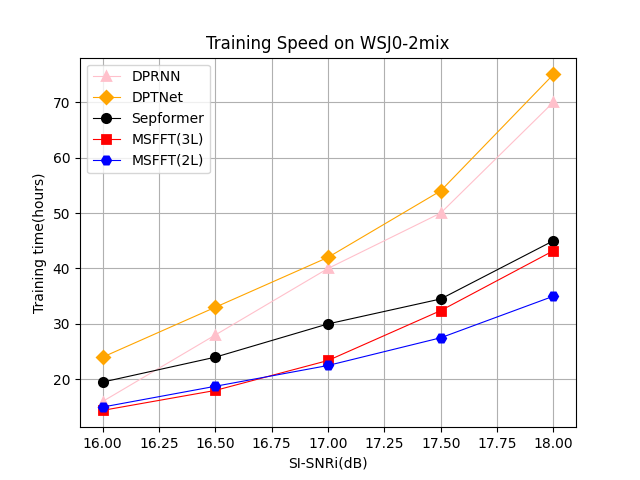}
    \caption{The training time for different speech separation methods.}
    \label{fig:9}
\end{figure}

Fig.~\ref{fig:9} plots the training time of different methods with respect to SI-SNRi. Obviously, both MSFFT-3P and MSFFT-2P have a faster training speed for obtaining the same SI-SNRi value than other dual-path based approaches. In the separation stage, the MSFFT-Net takes about 5 seconds to separate a mixed speech with duration of 5$s$, indicating relatively high efficiency of our proposed method in the inference stage. A Pytorch implementation with the demo page can be found at "http://isiplab.ahu.edu.cn/MSFFT/index.html".

\section{Discussion}
\label{sec:discussion}
The problem of speech separation comes from the cocktail party problem, where the goal of speech separation is to separate the main speaker's voice from the interference of other voices and noise in addition to the main speaker in the recorded audio signal. However, this task has proved to be easy for the human auditory system, but much more difficult for machines. Tasnet is an effective deep learning method for dealing with speech separation tasks, which consists of three modules: encoder, separation network and decoder. In this paper, we propose a novel separation network called multi-scale feature fusion transformer network (MSFFT-Net). We find that by fusing features from speech sequences at different scales, the features can be refined and enriched, which will improve the accuracy of mask estimation.

Then, we confirmed the effectiveness and generality of our proposed network through comparative experiments. Experiments show that our proposed networks achieve better results than the original dual-path models on the relevant datasets. 

However, the network that we introduce still has several shortcomings. Firstly, the parameter of our model is relatively large in comparison with other lightweight models such as DPRNN, GALR and Sandglasset. Secondly, the memory usage of our model is much heavier than DPRNN——our best model has a memory usage of almost 32GB.

In summary, our model achieves both outstanding performance and high efficiency. Moreover, our model uses the idea of multi-scale feature fusion, and although there have been previous studies using this idea for modeling, they have not reached SOTA. Thus, our work enriches the research in the field of speech separation using multi-scale feature fusion methods, which is an important guide for similar studies in the future.

\section{Conclusions}
\label{sec:conclusion}
In this paper, we introduce a multi-scale feature fusion transformer network (MSFFT-Net) based on the dual-path structure for end-to-end single-channel speech separation. According to previous research and experience, multi-scale feature fusion method can enrich feature information. Therefore, we introduce this idea in the speech separation network, after upsampling and downsampling operations to capture and fuse the speech sequence features at different scales. Experiments on relevant datasets also suggest the effectiveness and superiority of our models. In the future, we have two main works. Firstly, this paper has already introduced two different multi-scale feature fusion structures and both of them achieve better results than the original network, so we would try to design other multi-scale feature fusion structures as our future work. Secondly, the model parameters and memory usage of our model need to be reduced in future studies. Considering that transformer itself has relatively a large scale, we might transfer our multi-scale feature fusion structure to other lightweight models.

\section*{Acknowledgement}
This research has been supported by the National Natural Science Foundation of China Joint Fund Key Project(U2003207), National Natural Science Foundation of China(61902064), Natural Science Foundation of Anhui Province(No.1908085MF209), Key Projects of Natural Science Foundation of Anhui Province Universities(No.KJ2018A0018)

\bibliographystyle{IEEEtran}
\bibliography{Reference}

\end{document}